\documentclass[9pt,twocolumn,twoside]{osajnl}

\journal{ol} 

\setboolean{shortarticle}{true}

\usepackage{amssymb,amsmath}
\usepackage{graphicx}
\usepackage{graphics}
\usepackage{epsfig}
\usepackage{eurosym}
\usepackage{color}
\usepackage[normalem]{ulem}
\usepackage[makeroom]{cancel}

\newcommand{\PT}{{\cal PT}}

\newcommand{\rtext}[1]{\textcolor{black}{{#1}}}

\title{A universal form of arrays with spectral singularities
}

\author[1]{Dmitry A. Zezyulin}
\author[2]{Vladimir V. Konotop}

\affil[1]{ITMO University, St. Petersburg 197101, Russia}
\affil[2]{Centro de F\'{\i}sica Te\'orica e Computacional, Faculdade de Ci\^encias,  
	Universidade de Lisboa, Campo Grande, Edif\'icio C8, Lisboa  1749-016, Portugal
	and Departamento de F\'{\i}sica, Faculdade de Ci\^encias, Campo Grande, Edif\'icio C8, Lisboa  1749-016, Portugal}
\affil[*]{dzezyulin@itmo.ru}

\begin{abstract}
An array of non-Hermitian optical waveguides   can operate as a laser or as a coherent perfect absorber, which corresponds to a spectral singularity  of the underlying discrete complex potential.
We show that all  lattice potentials 
with spectral singularities are characterized by the universal form of the gain-and-loss distribution. Using this result  we   systematically construct   potentials characterized by   several spectral singularities at arbitrary wavelengths, as well as  potentials with second-order spectral singularities in their spectra. Higher-order spectral singularities demonstrate a  greatly enhanced response to incident beams resulting in the excitation of high-intensity lasing modes. 

\end{abstract}

\setboolean{displaycopyright}{true}

\begin{document}

\maketitle

Properly tuned complex potentials can totally absorb incident  electromagnetic radiation or emit radiation in the absence of incident waves, i.e., operate as coherent perfect absorbers (CPAs) or as lasers.  While the conceptual possibility of these phenomena was established long ago \cite{SSphysics_1,SSphysics_3,SSphysics_5}, they continue to attract growing attention  \cite{review,review_Rosanov} which is inspired by uncovering  the link between   perfect absorption and lasing and  the concept of spectral singularities (SSs)  
\cite{ScarfII,Mostafazadeh2009,Stone,Longhi}, as well as   by   the experimental observations of these phenomena~\cite{CaoScience,Wong16}.  Practical realization of  devices  characterized by   perfect absorption or  lasing   at the prescribed wavelength   therefore requires to design complex potentials with SSs in their spectrum. This problem is sufficiently well-studied in  one-dimensional continuous settings, where potentials with SSs can be constructed algorithmically \cite{Mostafazadeh2014_1,Mostafazadeh2014_2,KLV}. Moreover, it has been shown that all such potentials  have a universal form~\cite{ZK2020}. 

Absorbing and lasing potentials can be implemented also in discrete systems. Discrete perfectly absorbing potentials have been realized experimentally in Bose-Einstein condensates~\cite{matter_absorber} and in acoustic systems~\cite{Rivet2018}. They  have also been discussed for various systems of discrete optics~\cite{Ramezani2014,Jin2016,Longhi2018,CPA-optics}. 
In the meantime,  information on SSs for discrete light is still scarce. In this Letter, we show that discrete complex potentials with SSs in their spectra
have a {\em universal form} (similarly to their continuous counterparts~\cite{ZK2020}). Using this understanding, we can systematically design  one-dimensional arrays with  one or several SSs at wavelengths given beforehand, as well as second-order SSs.

We are interested in laser or CPA solutions of a waveguide array described by the dimensionless equation
\begin{eqnarray}
\label{SE_z}
id{q}_n/dz+\varkappa(q_{n-1}+q_{n+1}) - \gamma_n q_n=0, 
\end{eqnarray}
where $z$ is the propagation coordinate, and a set of complex parameters $\gamma_n$ is referred to as a  discrete potential. 
Considering only localized   potentials, i.e., such that $\lim_{n\to\pm\infty}\gamma_n= 0$,  we define an {\em SS-solution} $q_n(z)=e^{ibz}u_n$, where $b$ is real, and $u_n$  has the asymptotic behavior
\begin{equation}
\label{SSdefinition}
\lim_{n\to\infty}u_n e^{-ik_1n} = \rho_+, \quad \lim_{n\to-\infty}u_n e^{+ik_1n} = \rho_-,
\end{equation}
where $\rho_\pm$ are nonzero constants. 
We say that the wavenumber $k_1\neq 0$ is a SS. 
Equation (\ref{SE_z}) is characterized by the dispersion relation $b(k)=2\varkappa \cos k$, and 
thus SS-solutions exist only for $|b|< 2\varkappa$. For $k_1>0$ ($k_1<0$) the solution satisfying (\ref{SSdefinition}) describes a lasing (CPA) mode with purely outgoing (incoming) wave boundary conditions. 
Let us introduce a complex {\em base function} $w_n$ depending on the discrete variable $n$ and  having the asymptotic behavior $\lim_{n\to\pm\infty}w_n=\mp k_1$ or, more rigorously, characterized by the convergence of the series $\left|\sum_{j=0}^{\pm \infty}(w_j\pm k_1)\right|<\infty$.
Now we specify the class of discrete   potentials $\gamma_n$ 
defined by 
\begin{eqnarray}
\label{eq:gamma}
\gamma_n=-b+\varkappa\left(e^{iw_n}+e^{-iw_{n+1}}\right).
\end{eqnarray}
Clearly, $\gamma_n$ can be viewed as a function of $k_1$, $\gamma_n=\gamma_n(k_1)$ (hereafter we omit the explicit dependence on $k_1$).
Using the asymptotic properties of $w_n$ and the dispersion relation $b(k)$, one verifies that, in accordance with the above requirement, the discrete potential  (\ref{eq:gamma}) is indeed spatially localized.

It is straightforward to verify that Eq. (\ref{SE_z}) with potential (\ref{eq:gamma}) has an SS-solution 
\begin{equation}
\label{ss}
q_n= \rho_0 e^{ibz}\left\{
\begin{array}{ll}
    \exp\left(i\sum_{j=n+1}^{0}w_j\right) & \mbox{for } n\leq -1,
     \\
     1& \mbox{for } n=0,
     \\
    \exp\left(
-i\sum_{j=1}^{n}w_j
\right)  &  \mbox{for } n \geq 1,
\end{array}
\right.
\end{equation} 
where we introduced amplitude $\rho_0$ related to $\rho_\pm$ as  
$\rho_-=\rho_0\exp[i\sum_{j=-\infty}^{0}(w_j-k_1)]$ and $\rho_+=\rho_0\exp[-i\sum_{j=1}^{\infty}(w_j+k_1)]$.
Thus  
having a  
potential in the form  (\ref{eq:gamma}) is {\em sufficient} to enable a SS.  Invoking the analogy with the continuous case~\cite{ZK2020}, it is natural to conjecture that having a  
potential in the form (\ref{eq:gamma}) is also 
\emph{necessary}  
for the  existence of a SS. Indeed, suppose that Eq.~(\ref{SE_z}) has a SS, and $q_n$ is the respective SS-solution. Then, if $q_n\neq 0$ for all $n$, we define the base function  as $w_n=-i\ln({q_{n-1}}/{q_n})$,
which readily yields $\gamma_n$ in the form (\ref{eq:gamma}). Note, that discrete  potentials similar to that in Eq.~(\ref{eq:gamma})  were previously considered  
for a disordered system of discrete scatterers \cite{Makris2017,Rotter}. 

The obtained necessary and sufficient conditions represent a discrete analogue of the recently established universal form for continuous   one-dimensional potentials with SSs which, in the present notations, has the form $\gamma_n= -w^2_n  -id  w_n/d  n + k_1^2$  \rtext{\cite{ZK2020,Horsley}} that is  the long-wavelength 
limit of  potential (\ref{eq:gamma}).  

To  
confirm the above predictions, we  employ the transfer matrix approach. Far from the localized potential,  any   scattering state $q_n$  is  a superposition  of incident and reflected waves, i.e., $q_n \approx e^{ibz}(A_Le^{ikn} + B_Le^{-ikn})$ for $n\to-\infty$, and $q_n \approx e^{ibz}(A_Re^{ikn} + B_Re^{-ikn})$ for $n\to+\infty$. Here $k\in(-\pi, \pi)$,
and subscripts L and R mean ``left'' and ``right''.   The $2\times 2$ transfer matrix $M(k)$ relates the left and right coefficients as $(A_R, B_R)^T = M(A_L, B_L)^T$, where $T$ means transposition. A SS  $k_1$ corresponds to a zero of the transfer matrix element: $M_{22}(k_1)=0$ and, respectively, to singularities of left and right reflection and transmission coefficients computed as $r_L = -M_{21}/M_{22}$, $r_R =  M_{12}/M_{22}$,  $t_R=t_L = 1/M_{22}$. While for applications the scattering data are typically considered  only for positive wavenumbers, we formally evaluate them for all $k\in (-\pi,\pi)$. This allows to   distinguish between lasing SSs corresponding to scattering resonances with $k_1>0$ and CPA 
SSs with $k_1<0$.

\begin{figure}
		\includegraphics[width=0.99\columnwidth]{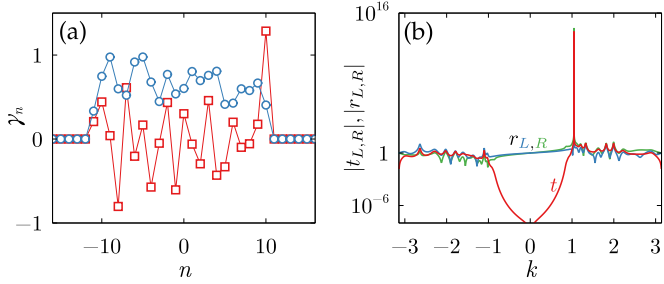}
	\caption{Example of a disordered potential with a lasing SS at $k_1=\pi/3$. (a) Real (circles) and imaginary (squares) part of $\gamma_n$. (b) Amplitudes of the transmission (red) and left and right reflection (blue and green) coefficients. Here and in all figures $\varkappa=1$. }
	\label{fig:random}
\end{figure}

 As an example, we consider discrete functions $w_n$ which are constant for all large positive and large negative $n$: specifically, we set $w_n=-k_1$ for $n>N$ and $w_n=k_1$ for $n<-N$. In the central region $-N\leq n\leq N$ we generate $w_n$ as a sequence of random (and generically complex) numbers. Then the  localized potential $\gamma_n$  is obtained with (\ref{eq:gamma}), and its transfer matrix elements and scattering coefficients are computed numerically. Example of such a ``disordered'' potential with a lasing spectral singularity is shown in Fig.~\ref{fig:random}, where the potential is obtained from a random complex sequence with $N=10$.
 
\begin{figure}
		\includegraphics[width=0.99\columnwidth]{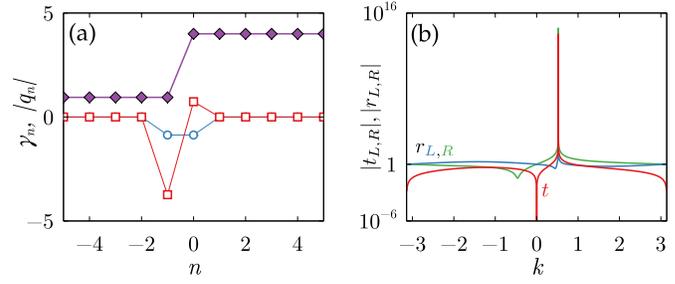}
	\caption{Example of lasing at strongly dominating overall dissipation. (a) Lasing mode at $k_1=\pi/6$ and $\rho_0=4$ (magenta diamonds)  shown together with the real (blue circles) and imaginary (red squares) parts of the discrete potential $\gamma_n$. (b) Amplitudes of the scattering
	coefficients.}
	\label{fig:absorption}
\end{figure}

A counterintuitive consequence of our result consists in the possibility to excite lasing modes in complex potentials characterized by arbitrarily strong overall absorption, i.e. with $\sum\mathrm{Im} \gamma_n$ being large negative. Lasing at relatively strong absorption is already seen in Fig.~\ref{fig:random}. To make this effect even more explicit,  we consider $w_n=-k_1\mathrm{sign}\,n$ for nonzero $n$ and $w_0=\arccos(i\Gamma)$, where $\Gamma$ is any real. The resulting potential is confined only to two sites, and the integral gain-and-loss  can be computed as $\sum\mathrm{Im} \gamma_n = 2\varkappa(\sin k_1 + \Gamma)$. Thus for large negative $\Gamma$ the system is subjected to strong overall   absorption. Nevertheless, for $k_1>0$ the potential always supports a lasing mode as shown in Fig.~\ref{fig:absorption}. The apparent  paradox is resolved after the inspection of the amplitude of the SS-solution $|q_n|$  which happens to be much larger in the active region than in the lossy domain. Conversely, for $k_1<0$ and large positive $\Gamma$ one can excite CPA-modes in spite of the overall strong gain. 

As an application of our result now we consider the construction of a potential with \emph{multiple SSs} at different wavevectors.  For obtaining a potential with {\em two} SSs, $k_1$ and $k_2$, it is  sufficient to find two different base functions $w_n^{(1)}$ and $w_n^{(2)}$, which after substitution in (\ref{eq:gamma}) result in the same  $\gamma_n$. Thus, we require:
\begin{equation}
2\cos k_1 - e^{iw_n^{(1)}} - e^{-iw_{n+1}^{(1)}} = 2\cos k_2 -  e^{iw_n^{(2)}} - e^{-iw_{n+1}^{(2)}}.
\end{equation}
Introducing new discrete functions  
\begin{equation}
\label{eq:chi}
\chi_n^{(j)} = \frac{1}{2}(e^{-iw_{n}^{(j)}} -e^{-iw_{n}^{(j+1)}}), \ 
\lim_{n\to \pm\infty}\chi_n^{(j)} = \frac{1}{2}(e^{\pm ik_j}  - e^{\pm ik_{j+1}}),
\end{equation}
we obtain a pair of    equations ($j=1,2$)
\begin{equation}
\label{eq:chi2}
\chi_n^{(1)} e^{2iw_n^{(j)}} +2(-1)^j \chi_n^{(1)}(c_{12}  - \chi^{(1)}_{n+1}) e^{iw_n^{(j)}}  + c_{12}  - \chi_{n+1}^{(1)} =0,
\end{equation}
where $c_{12} = \cos k_1 - \cos k_2$.
Equation (\ref{eq:chi2}) can be viewed as a quadratic equation with respect to $\exp(iw_n^{(j)})$.
Thus, for any $k_1$ and $k_2$ given beforehand, one can take any discrete function $\chi_n^{(1)}$ which satisfies asymptotic behavior from Eq.~(\ref{eq:chi}), and    use any of Eqs. (\ref{eq:chi2}) to compute $\exp(iw_n^{(j)})$ and subsequently to  recover the potential $\gamma_n$ using the obtained base function $w_n^{(j)}$. \rtext{In  particular, at $k_1=-k_2$ the obtained potential operates as  a CPA-laser, which is free from additional requirements, like for instance, $\PT$ symmetry  used in previous studies \cite{Longhi,Wong16,Laser-Absorber}.}

The freedom in the choice of the generating sequence $\chi_n^{(1)}$ allows one to use the obtained result also for construction of potentials with  {\em second-order} SSs, i.e. SSs corresponding to   second-order zeros of   $M_{22}(k)$.  The main idea of this construction can be formulated as a controlled ``colliding'' two first-order SSs. To this end, we consider a situation when $k_1$ is fixed and $k_2$ approaches $k_1$, which is achieved by adjusting $\chi_n^{(1)}$ viewed as a function of the  wavenumbers, i.e., $\chi_n^{(1)} = \chi_n^{(1)}(k_1, k_2)$. The second-order SS corresponds to the limit $k_2\to k_1$, where the two SSs collide and the respective SS-solutions ``merge''. We  therefore require $\lim_{k_2\to k_1} \chi_n^{(1)}=0$ for each $n$ and hence   $\lim_{k_2\to k_1}(\cos k_1 - \cos k_2- \chi_{n+1}^{(1)})=0$. Then using the L'H\^{o}pital's rule, from   (\ref{eq:chi}) we compute the limit
\begin{eqnarray}
\exp(2i\tilde{w}_n) := \lim_{k_2\to k_1 }\exp(2i w_n^{(1)}) = \lim_{k_2\to k_1 }\exp(2i w_n^{(2)}) = 
	\nonumber\\= \lim_{k_2\to k_1}\frac{\cos k_1-\cos k_2 -\chi_{n+1}^{(1)}}{-\chi_n^{(1)}} = \lim_{k_2\to k_1}\frac{\partial_{k_2}\chi_{n+1}^{(1)}-\sin k_2}{\partial_{k_2}\chi_n^{(1)}},
	\label{eq:2nd}
\end{eqnarray}
where we introduced the limiting base function $\tilde{w}_n$ and $\partial_{k_2}$ is a partial derivative with respect to $k_2$. 
If the latter limit exists, it can be used to determine  $\tilde{w}_n$ which, after substitution in (\ref{eq:gamma}), yields  a lattice potential with a second-order SS $k_1$.

\begin{figure}
		\includegraphics[width=0.99\columnwidth]{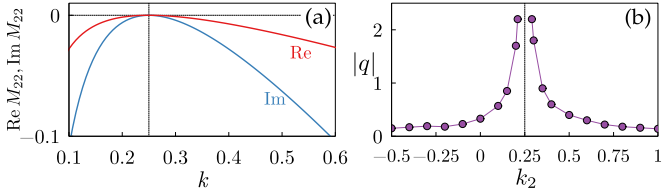}
	\caption{(a) Real and imaginary plots of the transfer matrix element $M_{22}(k)$ in the vicinity of the second-order SS at $k_{1,2}=0.25$. (b) Amplitude of the excited lasing mode at SS $k_1=0.25$ as the position of the second SS $k_2$ changes. Thin vertical line corresponds to $k_2=k_1$, where the two SSs collide and form a second-order SS, see Fig.~\ref{fig:2ndorder}.}
	\label{fig:M22}
\end{figure}

An example illustrating the described algorithm is given by $\chi_n^{(1)} = [\cos k_1-\cos k_2 \pm i  (\sin k_1-\sin k_2)]/2$ for  $\pm n\geq 1$. The resulting potential is zero for all $n$ except for  
\begin{equation}
\label{eq:simple}
\gamma_{-2}=-\frac{\varkappa}{2}(e^{-ik_1} + e^{-ik_2} \pm R),\,\,  \gamma_{-1}=-\frac{\varkappa}{2}(e^{-ik_1} + e^{-ik_2} \mp R),
\end{equation}
with $R=\sqrt{(e^{-ik_1} - e^{-ik_2})^2-4}$, and the  upper and lower signs correspond to two partner potentials. Using  above expression for $\chi_n^{(1)}$ together with (\ref{eq:2nd}), we obtain a pair of potentials with a second-order SS $\tilde{\gamma}_{-2}=-\varkappa(e^{-ik_1}  \pm i)$, $\tilde{\gamma}_{-1}=-\varkappa(e^{-ik_1}  \mp i)$. In retrospect, we could obtain this potential by sending $k_2$ to $k_1$ in (\ref{eq:simple}).  To confirm that the obtained SS is indeed of the second order, we plot real and imaginary parts of the transfer matrix element $M_{22}(k)$ in Fig.~\ref{fig:M22}(a), where we observe that its real and imaginary parts do not change sign in the vicinity of the SS, i.e., $k_1$ is indeed a  zero of the second order: $M_{22}(k_1)=dM_{22}(k_1)/dk=0$.

\begin{figure}
		\includegraphics[width=0.99\columnwidth]{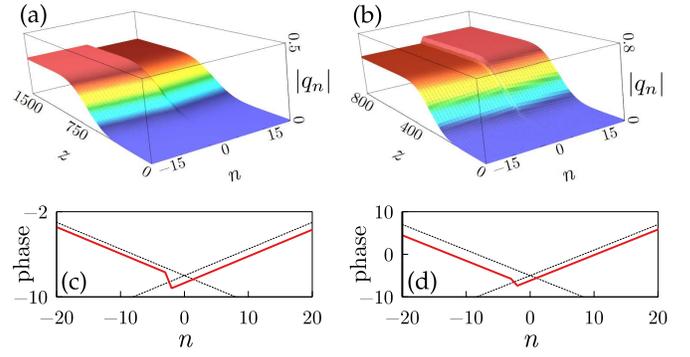}
	\caption{Excitation of lasing modes corresponding to SS $k_1=0.25$ (a) and $k_2=0.6$ (b) for potential (\ref{eq:simple}) by a small-amplitude left-incident beam   with wavevector $\nu=k_1$ and $\nu=k_2$;  (c,d) show unwrapped radian argument of the fields in the end of the shown propagation (red lines). Thin black lines have slopes  $\pm k_1$ (in c) and  $\pm k_2$ (in d) and are plotted for reference. 
	}
	\label{fig:2ss}
\end{figure}

Having in hands a potential with several SSs it is natural to question whether it is possible to  excite selectively a lasing or a CPA mode corresponding to a given SS. Here we  demonstrate that different lasing modes can be generated   by   probing   the potential with a properly designed  incident  wavepacket having sufficiently narrow spectral width.  In the beginning  of the propagation, the latter has the form of a localized quasi-monochromatic beam $Q_n e^{i\nu n}$, where $Q_n$ is a slowly decaying at $n\to\pm\infty$ envelope centered far to the left of the potential, and $\nu>0$ tunes the angle of  incidence. In Fig.~\ref{fig:2ss}(a) and (b) we show the result of the simulations of   beam propagation when the  input wavevector $\nu$ is equal to $k_1$ and $k_2$, respectively. In either case we observe the excitation of a lasing mode whose amplitude is about three orders of magnitude larger  than the maximal amplitude of the input beam (the latter was $\max |Q_n| =10^{-3}$ and is therefore indistinguishably  small in the scale  the plots). The difference between the two  emerging   lasing modes  is the best visible from their output phase distributions plotted in Fig.~\ref{fig:2ss}(c,d). In both cases the phase distributions are expected V-shaped~\cite{CPA-optics}, but the slopes of the decreasing and increasing segments are different in each mode and are determined by the wavenumber $\nu$ of the incident beam.  Thus the two obtained modes are totally distinct. 

\begin{figure}
		\includegraphics[width=0.99\columnwidth]{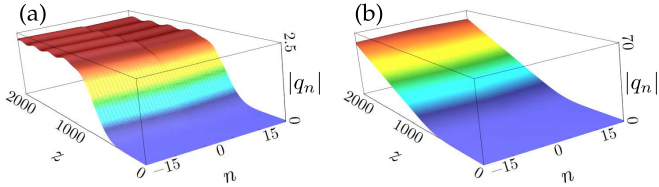}
	\caption{(a) Excitation of lasing mode with SS $k_1=0.25$ when the position of the second SS is set to $k_2=0.29$.  (b)  Propagation in the presence of  the second-order SS $k_1=k_2=0.25$. 
	Notice the different scales of amplitudes in panels (a) and (b).
	}
	\label{fig:2ndorder}
\end{figure}

To explore the collision of two SSs, we fix $k_1$ and study how the lasing mode corresponding to SS $k_1$ behaves as the second SS $k_2$ is driven to  approach $k_1$.  The result plotted in Fig.~\ref{fig:M22}(b) shows that in the vicinity of the collision the output lasing amplitude is enhanced dramatically [compare also Fig.~\ref{fig:2ss}(a) and Fig.~\ref{fig:2ndorder}(a)]. 
Exactly at the point of second order SS, we observe that the output amplitude grows along the propagation distance as shown in Fig.~\ref{fig:2ndorder}(b). \rtext{Such an enhanced response can be explained by difference in the interference patterns of the scattered waves. Indeed, considering superposition of two transmitted waves  (similar arguments are valid for reflected waves) with wavenumbers close to a SS (say, to $k_1$) $k_\pm=k_1\pm \delta k$ at a given site $n$, in the leading order one has $r(k_+)e^{ik_+n}+r(k_-)e^{ik_-n}\approx [2in M_{12}/M_{22}^\prime] e^{ik_1n} $ (prime stands for the derivative in the point $k_1$) in the case of a simple SS, and $r(k_+)e^{ik_+n}+r(k_-)e^{ik_-n}\approx [4M_{12}/M_{22}^{\prime\prime}](\delta k_1)^{-2} e^{ik_1n}$. This is similar to the enhanced scattering of beams in continuous systems~\cite{KLV} where it was shown that excitation of a given intensity output in scattering  process by a potential with a SS requires the less energy input signal the higher the order of a SS. Meantime, by exploring potentials different from (\ref{eq:simple}) but also featuring second order SSs we have found that enhancement may show more sophisticated dynamics associated with excitation of several modes (the result requiring further more delicate analysis).} 

Next, proceeding to the construction of a potential with {\em three} SSs: $k_{1,2,3}$, we have to find  three different base functions $w_n^{(j)}$ ($j=1,2,3$) yielding the same $\gamma_n$. Thus, now we have to find   two discrete functions  $\chi_n^{(1)}$ and $\chi_n^{(2)}$ for which equations
\begin{subequations}
\begin{eqnarray}
\label{eq:3ss1}
\chi_n^{(1)}e^{2i w_n^{(2)}} + 2\chi_n^{(1)}(c_{12} - \chi_{n+1}^{(1)})e^{i w_n^{(2)}}  + c_{12} -\chi^{(1)}_{n+1} =0,\\
\label{eq:3ss2}
\chi_n^{(2)}e^{2i w_n^{(2)}} - 2\chi_n^{(2)}(c_{23} - \chi_{n+1}^{(2)})e^{i w_n^{(2)}}  + c_{23} -\chi^{(2)}_{n+1} =0,
\end{eqnarray}
\end{subequations}
where $c_{ij}=\cos k_i - \cos k_j$, are satisfied simultaneously. If two such functions are found, then the base function $w_n^{(2)}$ found from either of these equations can be used to recover    $\gamma_n$.

\begin{figure}
		\includegraphics[width=0.9\columnwidth]{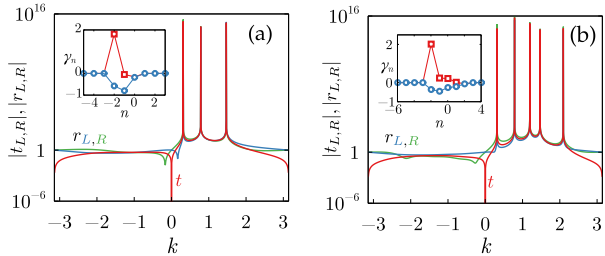}
	\caption{Potentials with three (a) and five (b) SSs. Main panels and insets show amplitudes of the  scattering coefficients and real (blue circles) and imaginary (red squares) parts of $\gamma_n$.	}
	\label{fig:3ss}
\end{figure}

To illustrate this algorithm, we consider 
\begin{equation}
\label{eq:3sschi}
\chi_n^{(j)} = 
(e^{\pm ik_j} - e^{\pm ik_{j+1}})/2, \quad\mbox{for $\pm n\geq 1$}, 
\end{equation}
with $\chi_0^{(1,2)}$ to be defined. Using $\chi_n^{(1)}$ we find $\exp({iw_n^{(2)}})$ from the quadratic equation (\ref{eq:3ss1}). In fact, it is sufficient to consider only $n=-1,0$, because for  all $n\leq -2$ and for all $n\geq 1$ values $\exp(iw_n^{(2)})$  are equal to  their limits $e^{\mp ik_2}$. Then we use the found quantities, to recover elements of $\chi_n^{(2)}$ from (\ref{eq:3ss2}):
\begin{subequations}
\begin{eqnarray}
\chi_0^{(2)} = \cos k_2 -\cos k_3 - {\chi_{-1}^{(2)}e^{2iw_{-1}^{(2)}}}
/(2\chi_{-1}^{(2)}e^{iw_{-1}^{(2)}} - 1)
\\
\label{eq:3ss_chi1}
\chi_1^{(2)} = \cos k_2 -\cos k_3 - {\chi_{0}^{(2)}e^{2iw_{0}^{(2)}}}
/{(2\chi_{0}^{(2)}e^{iw_{0}^{(2)}} - 1)}.
\end{eqnarray}
\end{subequations}
For the procedure to be consistent, the r.h.s. of (\ref{eq:3ss_chi1}) must coincide with the value  $\chi_1^{(2)}$ given by (\ref{eq:3sschi}). This leads to the closed-form equation, where 
the role of an unknown is played by  so-far unspecified  $\chi_{0}^{(1)}$. 
Since analytical solution of the  latter equation
is   complicated, we solve it numerically. In Fig.~\ref{fig:3ss}(a,b) we illustrate   a  potential which lases at three different wavenumbers.
Unlike in the case of two different SSs, now we do not have much freedom in the choice of $\chi_n^{(j)}$. Therefore  it is not straightforward to use this approach to create a third-order SS.

Generalizing the above ideas, we can construct potentials with practically any number $N\geq 3$  of SSs. To this end, we use substitution similar to (\ref{eq:3sschi}) but with $N-2$ unknown values $\chi_n^{(1)}$ at $n=0, 1, \ldots, N-2$. In the end,  we arrive at   $N-2$   equations  with respect to $N-2$ complex unknowns $\chi_0^{(1)}, \chi_1^{(1)}, \ldots, \chi_{N-2}^{(1)}$. The latter system can be   solved numerically. For example, in Fig.~\ref{fig:3ss}(c,d) we illustrate a potential with   \emph{five} different SSs.


To conclude, we have shown that discrete complex potentials allowing for lasing or perfectly absorbing solutions have a universal form. This fact enables systematic construction of potentials with spectral singularities at arbitrary wavelengths, including cases of multiple and higher-order spectral singularities. We also have shown that higher-order spectral singularities, compared with the simple ones, greatly enhance the system response, allowing excitation of high-intensity beams by incident beams of  very weak intensity. \rtext{Our results on construction of multiple spectral singularities can be  directly generalized to continuous potentials \cite{multipleSS}. Other possible extensions include studies of  the role of nonlinearities   of either medium~\cite{CPA-optics} or of scattering potentials~\cite{nonlin} for emergence of multiple and higher-order spectral singularities. Finally, it can be of interest to elucidate the eventual  dispersive \cite{dispersive} or chiral \cite{EP-CPA}  properties of  coherent perfect absorbers associated with   discrete spectral singularities.
}

\medskip

\noindent\textbf{Funding.}   Russian Foundation for Basic Research (RFBR) project No. 19-02-00193;
 Portuguese Foundation for Science and Technology (FCT)   Contract no. UIDB/00618/2020. 

\medskip

\noindent\textbf{Disclosures.} The authors declare no conflicts of interest.

\newpage

\section*{Full References}

\end{document}